\documentclass[12pt, preprint]{aastex}
\newcommand{\myemail}{astrosriram@yahoo.co.in}
\newcommand{\grs}{GRS~1915+105}
\newcommand{\x}{Cygnus~X-3}
\newcommand{\Au}{4U~1630-47}
\newcommand{\xtj}{XTE~J1550-564}
\shorttitle{Hard X-ray time lag: ~\xtj~ \& \Au}
\shortauthors{Sriram et al.}
\begin{document}
\title{Anti-correlated hard X-ray time lags in Galactic black hole sources}
\author{K. Sriram\altaffilmark{1}}
\affil{Department of Astronomy, Osmania University, Hyderabad-500007, India}
\author{V. K. Agrawal}
\affil{Tata Institute of Fundamental Research, Mumbai-400005. India}
\author{Jayant K. Pendharkar} 
\affil{Department of Astronomy, Osmania University, Hyderabad-500007, India}
\author{A. R. Rao}
\affil{Tata Institute of Fundamental Research, Mumbai-400005. India}
 
\altaffiltext{1}{e-mail: \myemail}
\begin{abstract}
We investigate the accretion disk geometry in Galactic black hole
sources by measuring the time delay between soft and hard X-ray emissions.
Similar to the recent discoveries  of anti-correlated hard X-ray time lags in \x~ and ~\grs~\citep{choudhury04apjl, choudhury05apj}, 
we find that the hard X-rays are anti-correlated with soft X-rays with a significant lag
in  another source:
XTE~J1550-564. We also find the existence of pivoting in the model independent X-ray spectrum during 
these observations. 
We investigate time-resolved X-ray spectral parameters  and
find that the variation in these parameters is consistent with the
idea of a truncated accretion disk.
The QPO frequency, which is a measure of the size of truncated accretion disk, too  changes 
indicating  that the geometric size of the hard X-ray
emitting region changes along with the spectral pivoting and soft X-ray flux. Similar kind of delay is also noticed in ~\Au. 
\end{abstract}
\section {Introduction}

Galactic black hole candidate sources provide an unique platform to study the
 behavior of ambient accreting material in intense gravitational fields. The spectral and timing analysis of accreting black hole 
sources gives information about the underlying physical phenomenon and dynamics of the 
accretion disk,  which in turn helps in unfolding
 the nature of the central black hole. The emission properties of the accreting black holes are often classified and constrained in terms of different spectral 
states \citep{esin97apj}. Timing analysis too plays a key role in uncovering the geometry of the disk, 
especially the study of Quasi Periodic Oscillations (QPO).
In a recent review,  \citet{mcclintock04} used both
 spectral and timing information to constrain the physics and geometry of the accretion disk. Canonically speaking the different spectral states are different permutation of 
 two spectral components i.e thermal component (soft photons),
presumably originating from a optically thick disk and Comptonized component (hard photons) 
which is thought to be the result of inverse Compton scattering of soft photons by 
high energy electrons  \citep{shapiro76apj, sunyaev80aa}.  The coupling of jet and the 
accretion disk in various spectral states of different galactic black hole systems 
\citep{fender05apss, fender06mnras} provided a new insight to the accretion disk geometry and
there are attempts to explain some parts of the observed X-ray spectrum as arising from 
jet emission \citep{vadawale01aa, markoff04apj, markoff05apj}. 
 
The bulk of the X-ray emission, particularly in the Very High State (or the Steep Power Law state),
however, is thought to be as due to Comptonization (see Done and Kubota 2006 for a comparison
of Comptonization and synchrotron models).  
The exact mechanism and geometry of this Comptonization
process, though, are still to be revealed. There are various models to account 
for the presence of the hard component in broad band spectrum of the galactic
 black hole candidates and the most favored one is hot quasi-spherical cloud inside
 a truncated disk \citep{zdziarski02apj}. 
 Among different theoretical models, Advection Dominated Accretion Flow (ADAF) \citep{narayan94apjl}  and Two Component Accretion Flow  (TCAF) \citep{chakrabarti96phr} predict that the disk truncation radius determines the segregated spectral states in Galactic black hole sources.

 Recently two sources, Cyg X-3 and GRS 1915+105 \citep{choudhury04apjl, choudhury05apj}
 provided support for the truncated accretion disk scenario on the basis of the detection
 of an anti-correlation between soft ($2-7$ keV) and hard X-ray photons ($20-50$ keV),
delayed by a few hundred seconds.
These sources also showed a 
pivoting behavior in the model independent spectrum.
 Here the hard lag implies that hard photons are lagging to the soft photons on timescales of 10s and 100s. The anticorrelated hard lag is defined as an opposite and delayed change in hard flux corresponding to change in soft flux. 
  The anti-correlated delay  \citep{choudhury05apj} in GRS $1915+105$ was discovered in the 
variability class $\chi$ when the source was in the spectral state C  \citep{belloni00aa}, 
which is also classified in the literature as the Steep Power Law (SPL) state \citep{mcclintock04} or 
the Very High State (VHS) or the Hard Intermediate State (HIMS).
 The wide band X-ray spectrum shows an additional spectral component (apart from the
canonical disk black body and a thermal Compton spectrum), which can be modeled as an additional
power-law \citep{rao00aal} or as due to Comptonization from electrons having non-thermal
power-law energy distribution \citep{zdziarski02apjl}. The recent discovery of a `jet-line' 
in the hardness-intensity diagram of black hole sources indicating the onset of
superluminal jet emission during a particular region of HIMS \citep{fender04mnras} underlines the
importance of understanding the detailed accretion disk geometry in such states.
In this perspective, using the hard X-ray delays and constraining the parameters
of a truncated accretion disk has the potential of unraveling the elusive disk-jet connection.

 Recently \citet{done06mnras} have made a detailed wide-band spectral 
fitting to   the VHS state of the source XTE J1550-564 during the 1998 outburst and have
obtained  different geometric configurations of the thermal disk and the corona. They
find that, compared to the High Soft State of the source, the thermal disk has either
reduced its size or changed its emission properties (or both) and the geometry of the
corona is constrained to be within a truncation radius. 

We have searched for anti-correlated hard X-ray delays 
in this source for confirming the robustness of truncated accretion disk paradigm.
\Au~ is another source which shows a behavior pattern
similar to ~\grs~ and ~\x~ and we have searched for delays in this source too. 

XTE J$1550-564$ is a well known microquasar with
an identified  optical companion star of late-type subgiant (G8 IV-K4 III).
The mass of the  black hole is
$10.0\pm1.5$ $M_{\odot}$ and the binary inclination is $72^{\circ}\pm 5^{\circ}$ \citep{orosz02apj}.
It was discovered by the {\it All Sky Monitor} (ASM) onboard the {\it Rossi X-ray Timing Explorer} (RXTE) on 1998 September 7 just after an  outburst which began on 1998 September 6 
\citep{smith98iau}. A few days later (1998 September 19),  a radio flare 
associated with a strong X-ray flare was detected, revealing
a relativistic jet when the X-ray source was in the 
very high state (VHS) \citep{hannikainen01apss}. 
A steady jet was discovered during the 2000 outburst but this time the source was
in the low-hard state \citep{corbel01apj}. Correlation studies between quasi periodic oscillations (QPOs) in the range of $\sim 0.08-22$ Hz
 and spectral parameters (multi-temperature black body disk and
 power-law component) indicate that the production of QPO is intimately tied to both disk and
 power-law components and is linked with the overall emission properties of the source \citep{sobczak00apj}. 
Detection of the high frequency QPOs ($\sim$ $185$ and $276$ Hz) \citep{remillard99apjl, remillard02apj} in ~\xtj~ and  ~\grs~($\sim$ $166$ Hz) \citep{belloni06mnras}
and the relationship of the 1 -- 15 Hz QPO with the spectral parameters
indicates an identical accretion disk geometry in these two sources \citep{markwardt99apj}.

Since its discovery by {\it Uhuru} in 1972, 4U $1630-47$, a recurrent X-ray transient  \citep{jones76apjl} located in the direction of the Galactic center, 
has shown all the different types of spectral states and Power Density Spectrum (PDS) 
like other black hole binaries and hence it is considered as a
strong black hole candidate source \citep{mcclintock04}. 
The light curves of 4U 1630-47 show different types of high amplitude variability, 
classified in four different classes \citep{tomsick05apj}
but the diversity 
of these variabilities is  less than that seen in the extremely variable source 
GRS 1915+105 \citep{belloni00aa}. On several occasions a very high disk temperature 
\citep{tomsick05apj} as well as polarized radio emission were seen 
in 4U $1630-47$ \citep{hjellming99apj}, indicating the presence of
relativistic jet emission, similar to the superluminal 
radio emitting jets seen GRS 1915+105. 

Here we report the discovery of  anti-correlated hard X-ray delay between soft 
and hard energy photons for both these sources. Paramount importance is given to know 
the exact variation in the physical parameters and hence unfolded spectra
are compared with the one from GRS 1915+105 obtained 
during the observation of delayed  hard X-ray emission. Spectral study is also supported
 by a   minute change in centroid frequency ($\sim$ 0.1 Hz) of the 
fundamental and the first harmonic of the Quasi Periodic Oscillations (QPO) in XTE J$1550-564$. 

\section {Data reduction and Analysis}
We have used data from observations using  Proportional Counter Array (PCA) \citep{jahoda06apjs} 
and High-Energy X-ray Timing Experiment (HEXTE) 
\citep{rothschild98apj} aboard the {\it RXTE} 
satellite to carry out a detailed temporal and spectral analysis. 
Done and Kubota  
\citep{done06mnras} have analyzed simultaneous ASCA and RXTE data on XTE J$1550-564$
and found evidence for an inner corona coupled energetically to the
disk. We have chosen the RXTE observations used in that work to look for 
anti-correlated delay. There were five pointed observations 
on three occasions, 1998 September 12, 23 and 1999 March 17 
(MJD $51068$, MJD $51079$ and MJD $51254$). The  X-ray spectrum from the  second data set 
shows strongly Comptonized VHS \citep{kubota04mnras}. We have used
standard 2 data and have followed all the procedures for  data filtering,
background and deadtime corrections (the data was obtained from all the
PCUs which were 'on').   
 HEXTE detectors
switch between background and source positions. During our observations
two sided rocking with 32 seconds rocking interval was selected. We used
FTOOLS command "hxtback" to separate background and source data. 
We created source and background lightcurves
using a binsize equal to the rocking interval (32 s) and 
applied the dead time correction to both the light curves.
For the timing analysis, we rebinned the
lightcurve by a factor of 4 (128 s), thus ensuring that the source and
background counts are averages of two rocking intervals.
For the present work, we use data only from HEXTE cluster A, which has
a better sensitivity than HEXTE cluster B.

 To take care of the calibration uncertainties,
 $0.5\%$ systematic errors are added to the PCA spectrum. We have obtained
 the Power Density Spectrum (PDS) for both the sources from generic bin mode and single bit mode covering $2-20$ keV and $20-50$ keV energy band with 1 ms bin size. 
\citet{tomsick05apj}
have presented  a detailed analysis of RXTE data during two years 
of X-ray activity of 4U$1630-47$. We have chosen $34$ observation from MJD $52790 - 52849$ when 4U$1630-47$ was in the outburst state and the data is not contaminated by 
nearby sources. Generally it was found that during these observations PCU 2 was `on' for a   maximum 
duration and
 hence we have used PCU 2  for obtaining the light curves and the spectra. 
The data reduction and analysis were done by using HEASOFT (V 5.3.1)\footnote{http://heasarc.gsfc.nasa.gov/lheasoft/RelNotes\_531.html}, which consists of mainly FTOOLS (V 5.3.1), XORONOS (V 5.21) and XSPEC (V11.3.1).

\section {Hard X-ray delay}
\begin{figure}[H]
\figurenum{1}
\epsscale{0.50}
\includegraphics[width=10cm,height=11cm,angle=270]{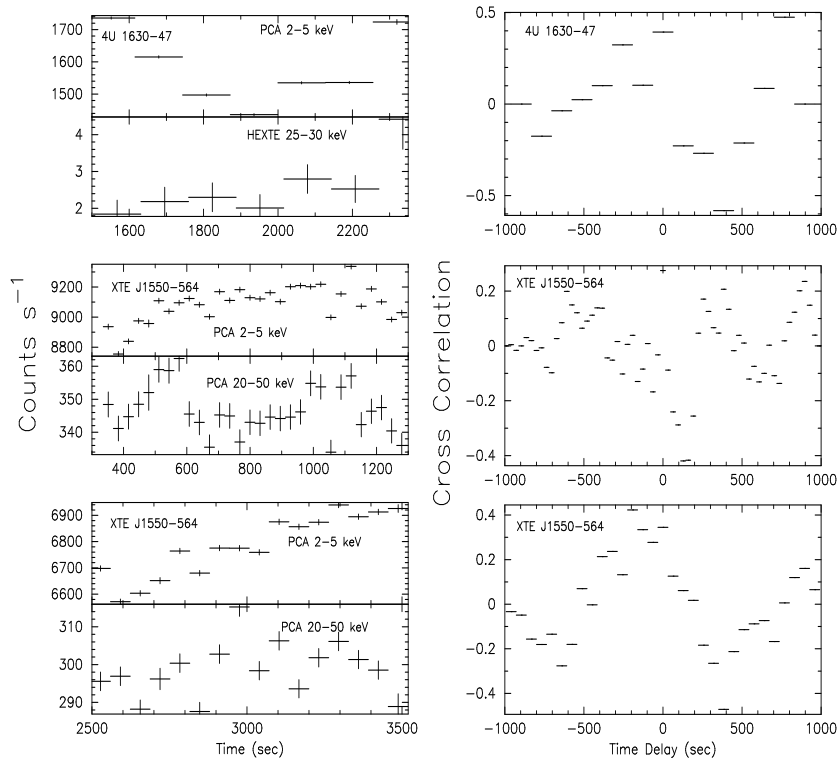}
\caption{The lightcurves and corresponding cross-correlation between the soft and hard X-ray flux in 4U $1630-47$ \& XTE $J1550-564$ \label{fig1} }
\end{figure}

During these observations, ~\xtj~ was in an outburst state and spectrally in the very high state or steep power law state (VHS or SPL). After obtaining the background subtracted lightcurve, we started with cross correlating the soft X-ray light curves ($2-5$ keV)
 and hard X-ray light curves ($20-50$ keV) using the {\it crosscor} program.
 The {\it crosscor} program performs cross correlation on two simultaneous
 time series by using a Fast Fourier Transform algorithm  
and the output is given as the cross correlation value as a function of
 time delay. The cross covariances are obtained by normalizing the cross correlations
 by dividing by the square root of the product of the number of good new bins of
 the pertinent lightcurves.  To calculate the observed delay and
 their uncertainties, we have fitted inverted Gaussian function to the anti-correlated
 hard delay part of the cross-correlation. Out of five observations two clearly
 show anti-correlated hard X-ray  delays of the order of a few hundred seconds (see Table \ref{tab1}). 
Similarly 4U$1630-47$ was also in an outburst state and spectrally mostly in SPL (steep power law) state, IS (intermediate
state) and
 spending a least amount of  time in TD (thermal dominated) state. Performing similar procedure
of timing analysis  we found that majority of them show a sharp positive correlation with no measurable
delay except for a few obsids which show some anti-correlated hard delay.
For one observation (ObsIds 80117-01-01-01, which is showing minute but considerable signature of delay), we have obtained the background subtracted light curves in different energy bands
 ($2-5$ keV, $5-10$ keV...$25-30$ keV) and found an anti-correlated hard
 delay ($\sim360$ sec) between $2-5$ keV and $25-30$ keV energy band, 
and this  particular observed delay was further supported by HEXTE analysis.
The cross-correlation values, as a function of delay, is shown in Figure \ref{fig1}, along
with the relevant light curves. To emphasize the reality of the delay seen
in 4U$1630-47$, analysis results from HEXTE observations are shown.
The observed delays are given in Table \ref{tab1}, along with the 90\% confidence errors (obtained by the
criterion of $\Delta \chi^2$ = 2.7, for an inverted Gaussian fit to the data).

\section {Spectral evolution}
Since the anti-correlated hard X-ray delay could cause pivoting pattern in
the spectrum in a single observation as observed in Cyg X-3 and GRS 1915+105
\citep{choudhury04apjl, choudhury05apj}, we divided the first (ObsId  30191-01-09-00)
and second (ObsId  30191-01-09-01) observations of XTE J$1550-564$ in two parts and extracted the spectra
covering $2-50$ keV  energy band. The observed spectra are shown in Figure \ref{fig2}.
  Inspection of the model
independent spectra for the first observation (Figure \ref{fig2}a), reveals
a sharp pivoting around $8-11$ keV. In the case of second observation a
marginal pivoting around 20 keV is observed. It is also noted that for the
second observation the spectra merge above 20 keV,  similar to that observed
in GRS 1915+105 during MJD 50729 \citep[see Figure 3]{choudhury05apj}. We argue that the
pivoting/marginal pivoting in the spectrum is due to hard lags in
the truncated accretion disk scenario. 
 \citet{koerding04aa} showed that the lag patterns in Cyg X-1 can be 
reproduced with a simple pivoting power law. 

\par To know the change in the spectral parameters during these observations,
 we have extracted the spectrum at
two different parts of the observation (initial and final $300$ seconds
of the light curve). 
We have used a multicomponent model which includes
disk blackbody \citep{makishima86apj} plus thermal Comptonization model
\citep[thcomp]{zdziarski96mnras} plus power-law to take care of high
energy non-thermal photons ({\it diskbb+thcomp+powlaw}) and a smeared edge
to mimic reflection component \citep{ebisawa94pasj}, along with a narrow Gaussian
line. Since the spectral resolution of PCA is not good enough to constrain
all the parameters, we have closely followed the spectral model derived from the
superior ASCA+RXTE simultaneous data 
\citep{done06mnras}. Further, the value of 
absorption column density, power-law index and Gaussian line energy 
were frozen (N$_{\rm H}$=$0.70 \times 10^{22} {\rm cm}^{-2}$, $\Gamma_{Pl}$=$2.2$ and E=6.5 keV).
The source was in strongly Comptonized very high state during these
observations typically showing a disk temperature $\sim$ $0.80$
keV, $\Gamma_{th}$$\sim$$2.32-2.41$ and kT$_{e}$$\sim$$10.50$ keV.
The unfolded spectra are shown in Figure 4 and the derived parameters are
given in Table
\ref{tab2}. We confirm that leaving the power-law index and absorption column density free, 
the values of other parameters are not changed (but less constrained).

\par For ~\Au,  we have extracted the spectra in the energy band 2.0 -
 50.0 keV at two different parts of the lightcurve corresponding to
 the observation 80117-01-01-01.  The model independent spectra show
 marginal pivoting  around $\sim 10$ keV (see Figure \ref{fig3}). We have
 fitted the spectra of these two parts with a model which includes disk
 blackbody plus thermal Comptonization model and found that this model
 gives unacceptable fits. Hence we unfolded the spectra using {\it
 "diskbb+Power-law"}, which gave reliable values as shown in Table
 \ref{tab4}. The change in disk temperature is quite low and may
 not be high enough to describe the marginal pivoting pattern in the
 spectrum. The spectral parameters indicates that the source is in TD
 (thermal dominated state), in which the disk extends close to the last
 stable orbit. We speculate that the delay may be attributed to marginal
 change in the power law index. \\

\par We have also  analyzed one of the observations (MJD 50480) of GRS
$1915+105$ which show a delay of the order of $\sim 1000$ s. To unfold
the spectrum we have used the same multicomponent model that was used for ~\xtj~
(see Figure \ref{fig5}). Three parameters are frozen, 
N$_{\rm H}$=$6.00 \times 10^{22} {\rm cm}^{-2}$
 \citep{belloni00aa}, power law index $\Gamma_{Pl}=2.0$
and Gaussian line energy = 6.5 keV. It can be seen from Table \ref{tab3}
that the both electron temperature and thcomp normalization has changed
between two parts of the observation, and cause a sharp pivoting point
at around $\sim 7$ keV \citep[see Fig 3]{choudhury05apj}.  
 
We have calculated the unabsorbed disk flux (soft X-ray flux) and unabsorbed thermal Comptonization flux (hard flux) 
for both ~\xtj~ and ~\grs~ observations (Table \ref{tab7}). In all the observations,
the soft X-ray flux is changing reflecting the degree of variability of
the accretion disk.

 It can be noticed from the derived spectral parameters that the 
quality of data is not sufficient to discern the changes in all the
parameters. To investigate the most dominant change in spectral parameters, we
have investigated the minimum set of parameters that definitely show a
change, as demanded by the data. For this purpose 
we fitted the part A (first part of the respective observation) and
 part B (second part of the respective observation)
 spectra simultaneously keeping all the parameters tied to spectral parameters obtained by fitting
only the part A spectrum. This resulted in very high reduced $\chi^2$
value (eg. $\chi^2/dof=436/185$ for ObsId 30191-01-09-00, see Table \ref{tab5}), suggesting a spectral change between
the two parts of the observation. Then thcomp normalization (N$_{th}$) of these two parts were
allowed to vary independently. The F-test values given in Table \ref{tab5}
suggest that fit improved drastically. Then we allowed two other
parameters to vary, diskbb normalization (N$_{diskbb}$) and kT$_{in}$,
one by one. The fit again improved significantly in this process. When we
continued this method for all the spectral parameters, no considerable
improvement in the fit was observed. Hence, F-test analysis suggest
that normalization and disk parameters vary between two parts (A and B)
of a single observation.

\section {Quasi Periodic Oscillations}
XTE J1550-564 and 4U 1630-47 show prominent QPOs in their Power
Density Spectra (PDS) \citep{remillard99apjl, tomsick05apj}. To unveil
the hidden physical phenomenon behind the anti-correlated hard X-ray 
delay, we have obtained the PDS of the same individual observation at
two ends (300s each; part A and B) of the corresponding lightcurve binned at 1 ms
in the energy range $2-20$ keV(B\_500us\_4A\_0\_49\_H) and $20-50$ keV
(SB\_125us\_50\_249\_1s) and normalized the output to squared fractional rms (with
the white noise subtracted). The spectral study of 4U $1630-47$ shows
that the source is in TD state which is  further supported by the absence of QPO
feature in the PDS whereas XTE J1550-564 shows strong signature of QPO
with one harmonic when the source is in strong VHS state. To quantify
the nature of QPO parameters, we fitted a powerlaw to the continuum and
two Lorentzian function to the QPO profile (see Table \ref{tab6}). In the
PDS of energy band $2-20$ keV, we found a harmonic which is not present
in the PDS of $20-50$ keV.

 The PDS of XTE J1550-564 for ObsId 30191-01-09-00 (Part A and B)
are shown in Figure \ref{fig6}, separately for 2 -- 20 keV (Figure 6a) and
20 -- 50 keV (Figure 6b). For clarity, part B data are shifted down by a 
factor of 2. The best fit models are shown as continuous lines and the
residuals (as a ratio of data to model) 
are shown at the two bottom panels of each figure, respectively
for part A and B. The fitted centroid frequencies are shown as vertical 
lines. Similarly, PDS for the same source for ObsId 30191-01-09-01 are
shown in Figure \ref{fig7}. Clear shifts are seen in the centroid as well as 
peak frequencies, particularly in the data for the second ObsId.

In both the
observations clear and consistent change in the centroid frequency of the
fundamental and the first harmonic of  QPOs are recognized.  The PDS of
$20-50$ keV energy band reveals that in one case the fundamental centroid
frequency shows no significant shift and in another case significant
amount of change is quite observable. In a single pointed observation the
centroid frequency is increasing in both the fundamental and harmonic
QPOs clearly suggesting that the inherent source size (Compton cloud)
is decreasing giving rise to geometrically and physically larger disk
\citep{chakrabarti00apjl}. The firm and undeniable change in
fundamental and harmonic centroid frequency demonstrates that there is
small but effective degree of variation in physical scenario of the
accretion disk and favors a truncated accretion disk. \\

\section {Discussion}
 Detection of anti-correlated hard delay in Cyg X-3 and GRS
1915+105 \citep{choudhury04apjl, choudhury05apj} suggested dynamical evidence of truncated accretion
disk. We anticipated the same physical scenario in two other sources
~\xtj~ and \Au.  The aim of this paper is to quantify the variation in
the spectral parameters responsible for the hard X-ray delay which in
turn constrains the geometry of the accretion disk.  Proper inspection
of the model independent spectrum indicates that, even a marginal
pivoting can account for the hard X-ray delay.

During our observations the sources ~\xtj~ and ~\grs~ were in the
VHS. Spectral analysis of these two sources  suggest that the 
soft and hard spectral components varied significantly, along with a 
strong indication of a change in the inner disk temperature.
We noticed a nominal change in the thcomp parameters in
both these sources. These changes are indirectly connected to the hot
electrons putative to the corona. 
In the case
of GRS 1915+105, the increase in the electron temperature is very  significant.

We have also calculated the unabsorbed bolometric flux for both disk
(soft flux) and thermal Comptonization components (hard flux) (shown
in Table \ref{tab7}). For both  the observation of ~\xtj~ we found
that whenever soft flux changes during the pivoting of the spectra,
an opposite change in hard flux occurs.  Similarly, we noticed that in
~\grs, the  soft flux  increases during the pivoting and an opposite
change in the hard flux is observed. This indirectly implies that there
is a change in either the geometry or physical properties of the accretion
disk and the corona. The time scales of delay observed in these two sources
suggest property of disk and corona changes on viscous time scales. The
observed delay is the readjustment time scale 
of the thermal disk as well as inner Comptonizing cloud. Since one
requires a truncated accretion disk to convert soft X-ray photons to
hard X-ray photons via Comptonization process and hence the extent of
the truncation radius decides the fate of the pivot point or change in
the respective flux.

The power density spectra of the source ~\xtj~ clearly shows a minute
change in the centroid frequency of the fundamental and the harmonic
QPOs. This small change in the centroid frequency of QPOs observed in  XTE
J$1550-564$ indicates the physical aberration of corona on a timescale
of $\sim$ 1000 s. There are various models \citep{titarchuk04apj,
chakrabarti00apjl} which explain the production of QPOs and these
models seem to converge to the fact that the origin of QPOs is inherent
to the compact corona region close to the black hole. In both the
observation of the source ~\xtj, the change in the centroid frequency is
strongly correlated to change in the soft and hard fluxes. In both the
observations, the centroid frequency is increasing along with the increase
in soft flux and decrease in the hard flux (see Table \ref{tab7}). In the
case of ~\grs, the hard flux is increasing along with a decrease in the soft
flux showing corresponding change in the QPO frequency. The proportional
change in the QPO centroid frequency with hard and soft fluxes  strengthens the
idea that during these hard lag observations there is a inward/outward
movement of the Compton cloud.

 The second observation of ~\xtj~ (ObsId 30191-01-09-01) shows the most
well determined delay  (375$\pm$13 s) and the most significant
shift in the QPO frequency. Let us assume that the truncation radius, R$_c$,  is at
10 Schwarzschild radii  (R$_{RS}$), consistent with the $\sim$  300 km 
derived for the disk-corona coupling radius \citep{done06mnras}.
The observed 5\% increase in the QPO frequency, f,  corresponds to
a 15\% decrease in
the truncation radius, by assuming a Lense-Thirring precession model
for the QPO generation \citep{done06mnras, stella99apj}
which predicts
f $\sim$ R$_c^{-1/3}$. This decrease of 15\% in R$_c$ is possibly due to an
increase in the disk accretion rate by $\sim$20\%, if R$_c$ varies as 
an inverse power $\beta$ of mass accretion rate through the disk
with $\beta \sim$  0.65 -- 0.85,  as  suggested by \citet{mchardy06nature}.
The increase in the 
disk luminosity would be about 35\% (scaling directly as the accretion rate and
inversely as R$_c$, as would be the case for a standard thin disk), which
is very close to the observed variation of 33\% (see Table 7). A corresponding
decrease in the Compton flux will occur after a delay corresponding to
the Compton cooling time scale for the plasma confined within R$_c$. 
By assuming similar properties of accretion disk as seen in
GRS~1915+105 during the inner-disk evaporation, a Compton cooling 
time scale of several hundred seconds \citep{chakrabarti00apjl}
can be derived, similar to the observed delay of 375 s.

On the other extreme, 4U 1630-47 shows significant anti-correlated
hard delay in thermal dominated state. We find that index of power law
component increases during the pivoting and at the same time soft flux
decreases. Hence, we suggest that pivoting is primarily due to decrease
in the soft disk flux and hardening of the power law component. Since
during the observation QPO was not observed we can not get any concrete
conclusions about physical changes in the accretion disk and corona.

 The five distinct properties viz. anti-correlated hard delay, change
 in QPO centroid frequency, model independent spectrum, systematic
 variation in thermal Comptonization parameters and disciplined change in the QPO frequency with fluxes, swings the pendulum
 towards the idea of truncated accretion disk and alleviate the
 possibility of non truncated models.

\section*{Acknowledgments} We thank the anonymous referee for constructive and critical comments. This research has made use of data obtained through
the HEASARC Online Service, provided by the NASA/GSFC, in support of NASA High
Energy Astrophysics Programs.

\clearpage
\begin{deluxetable}{llllll}
\tablecolumns{5}
\tablewidth{0pc}
\tabletypesize{\footnotesize}
\tablecaption{Details of the anti-correlated delays of the hard X-rays with respect to soft X-rays.\label{tab1} }
\tablehead{
\colhead{source} & \colhead{soft vs hard energy range} & \colhead{ObsId} & \colhead{bin (s)}&\colhead{ Delay (s)} &\colhead{correlation coefficient} }
\startdata
4U 1630-47 & 2-5 keV vs 25-30 keV & 80117-01-01-01 & 128 & $363\pm48$ & -0.58 \\
XTE J1550-564 & 2-5 keV vs 20-50 keV & 30191-01-09-00 & 32 & $132\pm9$ & -0.41\\
XTE J1550-564 & 2-5 keV vs 20-50 keV & 30191-01-09-01 & 64 & $376\pm14$ & -0.47\\
\enddata
\end{deluxetable}

\begin{deluxetable}{llllll}
\tablecolumns{8}
\tablewidth{0pc}
\tabletypesize{\footnotesize}
\tablecaption{Spectral parameters of XTE J$1550-564$ in an individual observations ($30191-01-09-00$ \& $30191-01-09-01$). \label{tab2} }
\tablehead{
\colhead{Part of observation} & \colhead{kT$_{in}\tablenotemark{a}$} & \colhead{$\Gamma_{th}\tablenotemark{b}$} & \colhead{ kT$_{e}\tablenotemark{c}$} & \colhead{ N$_{th}\tablenotemark{d}$ }& \colhead {$\chi^{2}/{\it dof}$}  \\
\colhead{}&\colhead{(keV)} & \colhead{} & \colhead{(keV)} & \colhead{} & \colhead{}  
}
\startdata
A & $0.82^{+0.03}_{-0.03}$ & $2.33^{+0.03}_{-0.03}$ & $10.61^{+1.20}_{-1.50}$ &
$4.24^{+0.60}_{-1.17}$ & $87.59/83$\\
B & $0.79^{+0.06}_{-0.02}$ & $2.41^{+0.03}_{-0.05}$ & $10.88^{+0.95}_{-1.68}$ &
$3.29^{+0.69}_{-1.36}$  &  $87.02/83$\\
\hline
\hline
A & $0.80^{+0.05}_{-0.07}$ & $2.31^{+0.03}_{-0.04}$ &  $10.55^{+1.77}_{-1.54}$&
$4.85^{+1.03}_{-1.70}$ &  $102.51/83$  \\
B & $0.80^{+0.02}_{-0.03}$  &  $2.32^{+0.03}_{-0.02}$ & $9.18^{+1.14}_{-0.81}$
& $2.70^{+0.42}_{-0.51}$ & $87.90/83$	 \\
\enddata
\tablenotetext{a} {inner disk temperature using "diskbb" model}
\tablenotetext{b} {Thermal Comptonization index}
\tablenotetext{c} {Electron temperature}
\tablenotetext{d} {Normalization of the "thcomp" model}
\end{deluxetable}

\begin{deluxetable}{lllllll}
\tablecolumns{10}
\tablewidth{0pc}
\tabletypesize{\scriptsize}
\tablecaption{Spectral parameters of GRS $1915+105$ in an individual observation (ObsId 20402-01-14-00) \label{tab3}}
\tablehead{
\colhead{Part of observation} & \colhead{kT$_{in}\tablenotemark{a}$} & \colhead{$\Gamma_{th}\tablenotemark{b}$} & \colhead{kT$_{e}\tablenotemark{c}$} &  \colhead{N$_{th}\tablenotemark{d}$} & \colhead{ $\chi^{2}/{\it dof}$} \\
\colhead{}&\colhead{(keV)}&\colhead{}&\colhead{(keV)}&\colhead{}&\colhead{}
}
\startdata
A& $0.84^{+0.02}_{-0.03}$  &  $2.25^{+0.03}_{-0.04}$ & $12.77^{+2.42}_{-1.85}$ &  $1.34^{+0.56}_{-0.40}$ &   $80.64/83$ \\
B& $0.82^{+0.02}_{-0.02}$  &  $2.22^{+0.02}_{-0.02}$  &  $19.00^{+4.05}_{-2.70}$ & $3.48^{+0.89}_{-1.40}$  & $78.48/83$ \\
\enddata
\tablenotetext{a} {Inner disk temperature using "diskbb" model}
\tablenotetext{b} {Thermal Comptonization index}
\tablenotetext{c} {Electron temperature}
\tablenotetext{d} {Normalization of the "thcomp" model}
\end{deluxetable}

\begin{deluxetable}{llllll}
\tablecolumns{6}
\tablewidth{0pc}
\tabletypesize{\footnotesize}
\tablecaption{Spectral parameters of 4U $1630-47$ in an individual observation. \label{tab4} }
\tablehead{ 
\colhead{Part of observation} & \colhead{N$_{\rm H}\tablenotemark{a}$} & \colhead{kT$_{in}\tablenotemark{b}$} & \colhead{$\Gamma_{PL}\tablenotemark{c}$} &  \colhead{$\chi^{2}/dof$} \\
\colhead{} & \colhead{(10$^{22}$ cm$^{-2}$)} & \colhead{(keV)} & \colhead{} & \colhead{}& \colhead{}
}
\startdata       
A & $11.44^{+0.55}_{-0.55}$ & $1.39^{+0.01}_{-0.01}$  & $3.41^{+0.11}_{-0.12}$ & $56.40/60$\\
B&$10.97^{+0.54}_{-0.51}$ & $1.37^{+0.01}_{-0.01}$ & $3.33^{+0.11}_{-0.12}$& $64.67/60$\\
\enddata
\tablenotetext{a} {Equivalent Hydrogen column density}
\tablenotetext{b} {Inner disk temperature using "diskbb" model}
\tablenotetext{c} {Power-law photon index}
\end{deluxetable}
\begin{deluxetable}{llllll}
\tablecolumns{6}
\tablewidth{0pc}
\tabletypesize{\footnotesize}
\tablecaption{ Results of simultaneous fitting of the different parts of a
single observations for the sources XTE J1550-564 and GRS 1915+105. 'NONE'
denotes that all the parameters were tied to the spectral parameters 
obtained from the first part of observation.
'ALL' denotes that all the spectral parameters were
allowed to vary independently.\label{tab5}  }
\tablehead{ 
\colhead{ObsId}&\colhead{parameters varied} & \colhead{$\chi^2$}& \colhead{DOF}&  \colhead{F-test probability} \\
}
\startdata       
(XTE J1550-564)&&&\\
30191-01-09-00& & &\\
&NONE & 436  & 185 & -\\
&{N$_{th}\tablenotemark{a}$} &285& 182 &  9.13 $\times$ 10$^{-17}$\\
&N$_{th}$+{N$_{diskbb}\tablenotemark{b}$} &245 &181 &1.744 $\times$ 10$^{-7}$\\
&N$_{th}$+N$_{diskbb}$+kT$_{in}\tablenotemark{c}$ &207 &180& 3.79 $\times$ 10$^{-8}$\\
&ALL & 195 &176  &3.18 $\times$ 10$^{-2}$\\
\hline
\hline
(XTE J1550-564)&&&\\
30191-01-09-01& & &\\
&NONE & 316 & 183 & -\\
&N$_{th}$  & 220  & 182  & 5.173 $\times$ 10$^{-16}$\\
&N$_{th}$+N$_{diskbb}$ &219 &181 &0.364\\
&N$_{th}$+N$_{diskbb}$+kT$_{in}$ &207 &180 &1.470 $\times$ 10$^{-3}$ \\ 
&ALL& 205  &176 &0.787\\
\hline
\hline
(GRS 1915+105)&&&\\
20402-01-14-00 & & &\\
& None&  976  & 151 &-\\
& N$_{th}$ & 819 &  150 & 3.046 $\times$ 10$^{-7}$\\
& N$_{th}$+N$_{diskbb}$ & 216 &149 & 5.605 $\times$ 10$^{-45}$\\
& N$_{th}$+N$_{diskbb}$+kT$_{in}$ &164 &148 & 1.843 $\times$ 10$^{-10}$\\
& ALL &150 & 144 &1.159 $\times$ 10$^{-2}$\\
\hline
\enddata
\tablenotetext{a} {Normalization of the "thcomp" model }
\tablenotetext{b} { Normalization of the "diskbb" model}
\tablenotetext{c} {Inner disk temperature using "diskbb" model}
\end{deluxetable}

\begin{deluxetable}{llllllll}
\tablecolumns{8}
\tablewidth{0pc}
\tabletypesize{\scriptsize}
\tablecaption{Details of the QPO parameters of XTE J$1550-564$ in the energy band of 2-20keV \& 20-50keV in an individual observation \label{tab6}}
\tablehead{
\colhead{ObsId} & \colhead{Energy} & \multicolumn{2}{c}{ Centroid frequency f (Hz)} & \multicolumn{2}{c}{FWHM (Hz)} & \multicolumn{2}{c}{RMS\%} \\
\colhead{} & \colhead{} & \colhead{Fundamental}&\colhead{Harmonic} & \colhead{Fundamental}&\colhead{Harmonic} & \colhead{Fundamental}&\colhead{Harmonic}
}
\startdata
$30191-01-09-00$ &Soft (2-20keV) & $4.32\pm0.02\tablenotemark{1}$ & $8.37\pm0.08$ & $0.52$ & $1.14$ & $1.64$ &  $0.18$    \\
                 &               & $4.38\pm0.02\tablenotemark{2}$ & $8.57\pm0.05$ & $0.41$ & $0.89$ & $1.36$ &  $0.18$\\
$30191-01-09-00$ &Hard (20-50keV)& $4.34\pm0.02$     & absent        & $0.48$ & absent & $2.73$ &  absent   \\
                 &               & $4.35\pm0.02$     & absent        & $0.38$ & absent & $2.81$ &  absent  \\
$30191-01-09-01$ &Soft (2-20keV) & $3.87\pm0.01$     & $7.64\pm0.04$ & $0.29$ &  $0.51$ & $1.41$ & $0.10$\\
                 &               & $4.07\pm0.01$     & $7.95\pm0.03$ & $0.36$ &  $0.53$ & $1.30$ & $0.10$\\
$30191-01-09-01$ &Hard (20-50keV)& $3.90\pm0.01$     & absent        & $0.30$ & absent & $2.42$ & absent   \\
                 &               & $4.05\pm0.05$     &absent          & $0.45$ & absent & $2.87$ & absent  \\
\hline
\enddata
\tablenotetext{1,2} {Represents different region of a single pointed observation }
\end{deluxetable}


\begin{deluxetable}{lllllll}
\tablecolumns{8}
\tablewidth{0pc}
\tabletypesize{\scriptsize}
\tablecaption{The values obtained from the simultaneous spectrum fitting for the respective sources. Three parameters (shown in table) were allowed to vary because of F-test results. The flux reported here are bolometric flux. For all the parameters, 90\% confidence errors are given. \label{tab7}}
\tablehead{
\colhead{Spectral parameter} &  \multicolumn{2}{c}{ 30191-01-09-00} & \multicolumn{2}{c}{30191-01-09-01} & \multicolumn{2}{c}{20402-01-14-00} \\
\colhead{} & \colhead{A}&\colhead{B} & \colhead{A}&\colhead{B} & \colhead{A}&\colhead{B} 
}
\startdata
{N$_{th}\tablenotemark{a}$}   &  $4.59\pm0.65$ &  $4.46\pm0.72$ & $3.33\pm0.62$ & $3.91\pm0.66$ & $3.16\pm1.45$&$3.02\pm1.35$ \\
{N$_{diskbb}\tablenotemark{b}$} & $2253\pm652$& $1600\pm476$  & $3500\pm626$  & $2557\pm882$  & $614\pm34$ & $824\pm92$\\
kT$_{in} \tablenotemark{c}$(keV)  & $ 0.80\pm0.02$ & $   0.87\pm0.04 $ & $ 0.73\pm0.01$ & $  0.78\pm0.02$ & $ 0.88\pm0.01$ & $0.79\pm0.02$\\
disk flux&  $20.9$&$21.9$ & $17.7$ & $23.5$ & $8.50$ & $6.61$ \\
{(10$^{-9}$ ergs cm$^{-2}$ s$^{-1}$)}&&&&&&\\
thcomp flux & 52.5 &46.5& 52.6&	42.3& 11.40&22.0\\
{(10$^{-9}$ ergs cm$^{-2}$ s$^{-1}$)}&&&&&&\\\\
 Delay (seconds)&  \multicolumn{1}{r}{$132\pm9$} && \multicolumn{1}{r}{$375\pm13$} &&\multicolumn{1}{r}{$704\pm35$} \\
{($\Delta$f/f)}\tablenotemark{d}\%  &    \multicolumn{1}{c}{1.39}  &  &         \multicolumn{1}{c}{5.16}  &&  \multicolumn{1}{c}{-10.03}\\
($\Delta$kT$_{in}$/kT$_{in}$)\% & \multicolumn{1}{c}{8.75}   &&       \multicolumn{1}{c}{6.85}	&&  \multicolumn{1}{c}{-10.23}\\
($\Delta$N$_{diskbb}/{\rm N}_{diskbb})\%$ &\multicolumn{1}{c}{4.78}&&\multicolumn{1}{c}{32.76}&&\multicolumn{1}{c}{-22.2} \\
($\Delta$N$_{th}$/N$_{th}$)\% & \multicolumn{1}{c}{-11.43}&&\multicolumn{1}{c}{-19.6}&&\multicolumn{1}{c}{92.98}\\
\hline
\enddata
\tablenotetext{a} {Normalization of the "thcomp" model }
\tablenotetext{b} { Normalization of the "diskbb" model}
\tablenotetext{c} {Inner disk temperature using "diskbb" model}
\tablenotetext{d} {f corresponds to centroid frequency of respective QPO }
\end{deluxetable}

\clearpage

\clearpage
{\begin{figure}
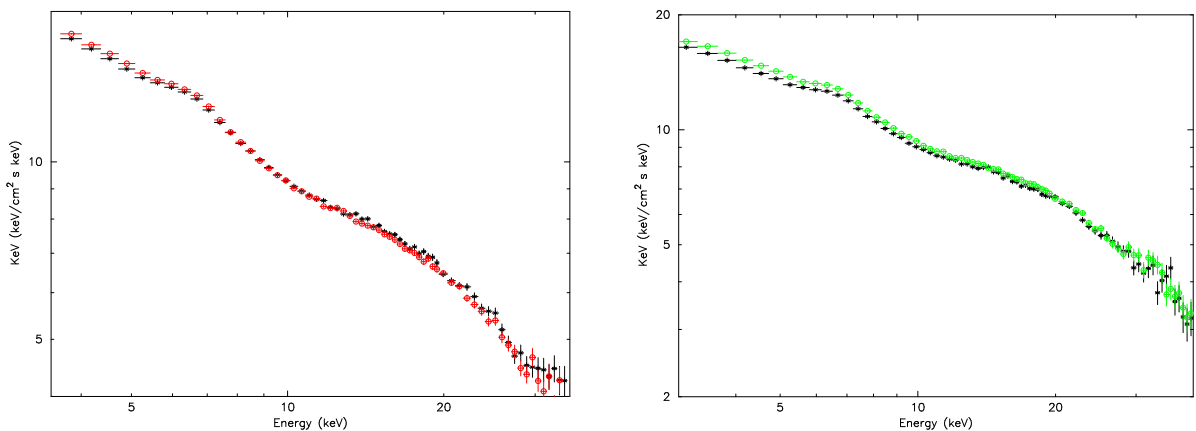

\figurenum{2}{
\begin{tabular}{c@{\hspace{2pc}}c}
\includegraphics[width=2.2in,angle=270]{f2a.eps} &
\includegraphics[width=2.2in,angle=270]{f2b.eps} \\
\end{tabular}}
\caption{Pivoting of the X-ray spectrum during the  observations of anti-correlated hard 
X-ray delays in  ~\xtj~ on two occasions. Left panel (Figure 2a, ObsId 30191-01-09-00) shows clear pivoting pattern around $\sim 8-11$ keV. Right panel (Figure 2b, ObsId 30191-01-09-01) shows change in the normalization around $\sim 10$ keV and merging of individual spectrum at higher energies. \label{fig2}  }
\end{figure}}

\begin{figure}
\figurenum{3}
\epsscale{0.50}
\includegraphics[width=5cm,height=8cm,angle=270]{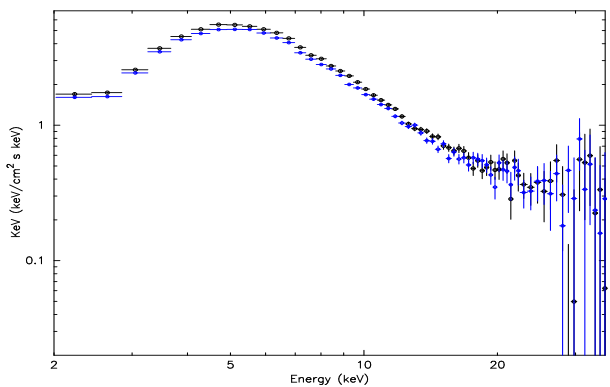}
\caption{Pivoting of the X-ray spectrum during the  observations of anti-correlated hard
X-ray delays in ~\Au. Change in the normalization is observed around $\sim 10$ keV. \label{fig3}}
\end{figure}

\clearpage
{\begin{figure}
\figurenum{4}{
\epsscale{0.50}
\begin{tabular}{c@{\hspace{2pc}}c}
\includegraphics[width=2.2in,angle=270]{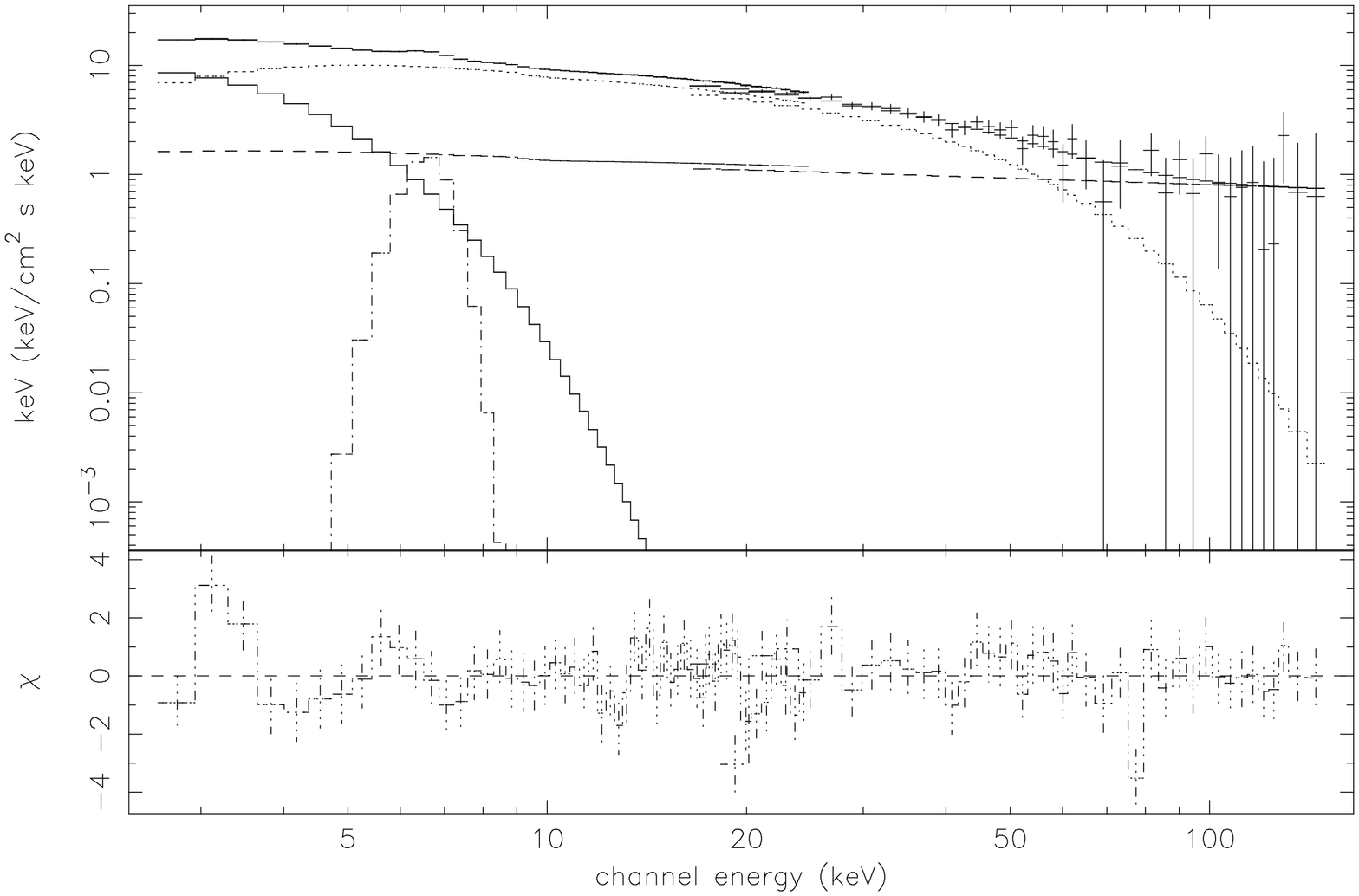} &
\includegraphics[width=2.2in,angle=270]{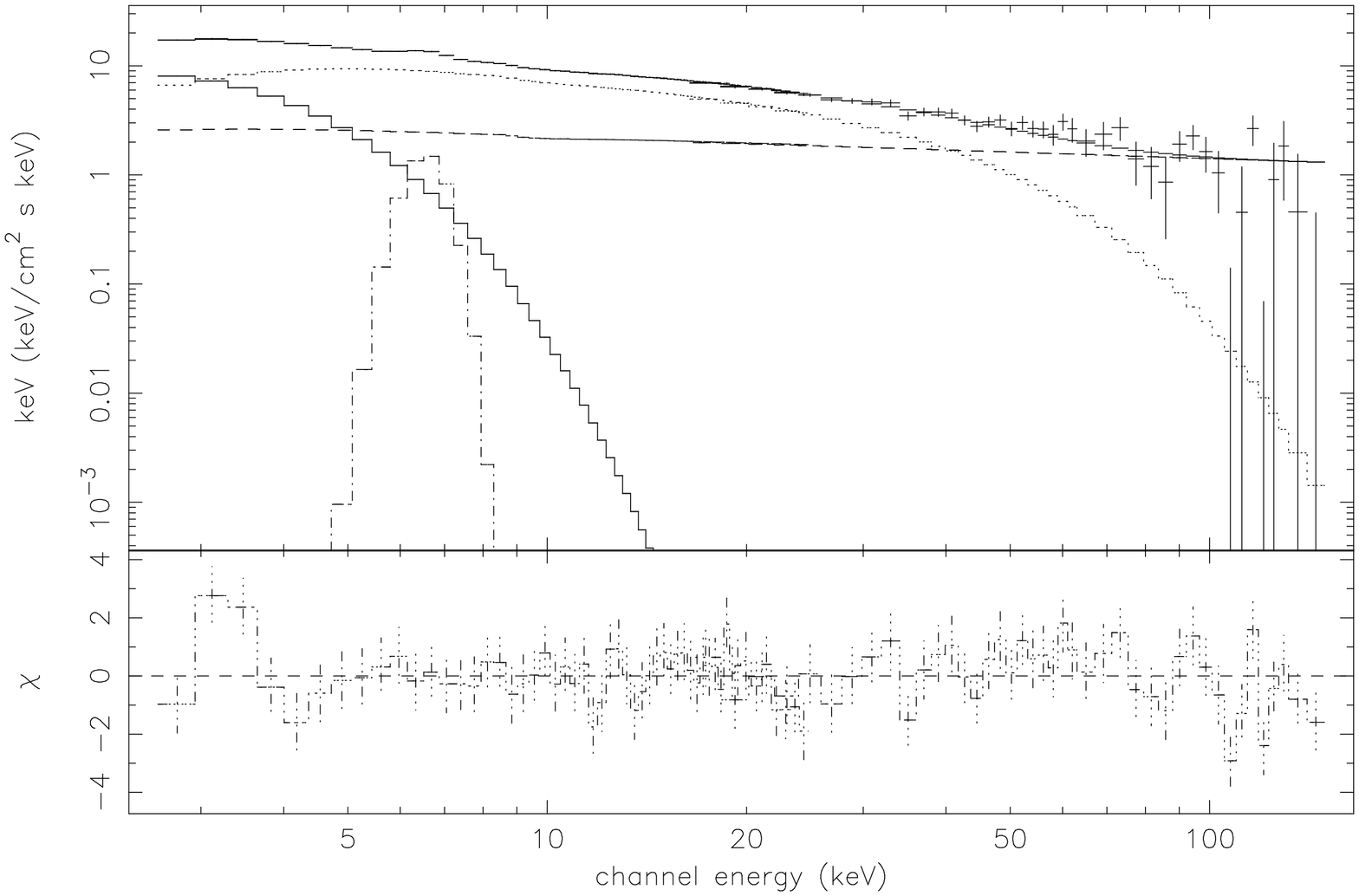} \\
\end{tabular}}
\caption{The unfolded spectrum for the source XTE J$1550-564$ for the different parts (left panel A \& right panel B, see Table \ref{tab2}) of a single observation (ObsId 30191-01-09-00). \label{fig4}}
\end{figure}}

{\begin{figure}
\figurenum{5}{
\begin{tabular}{c@{\hspace{2pc}}c}
\includegraphics[width=2.2in,angle=270]{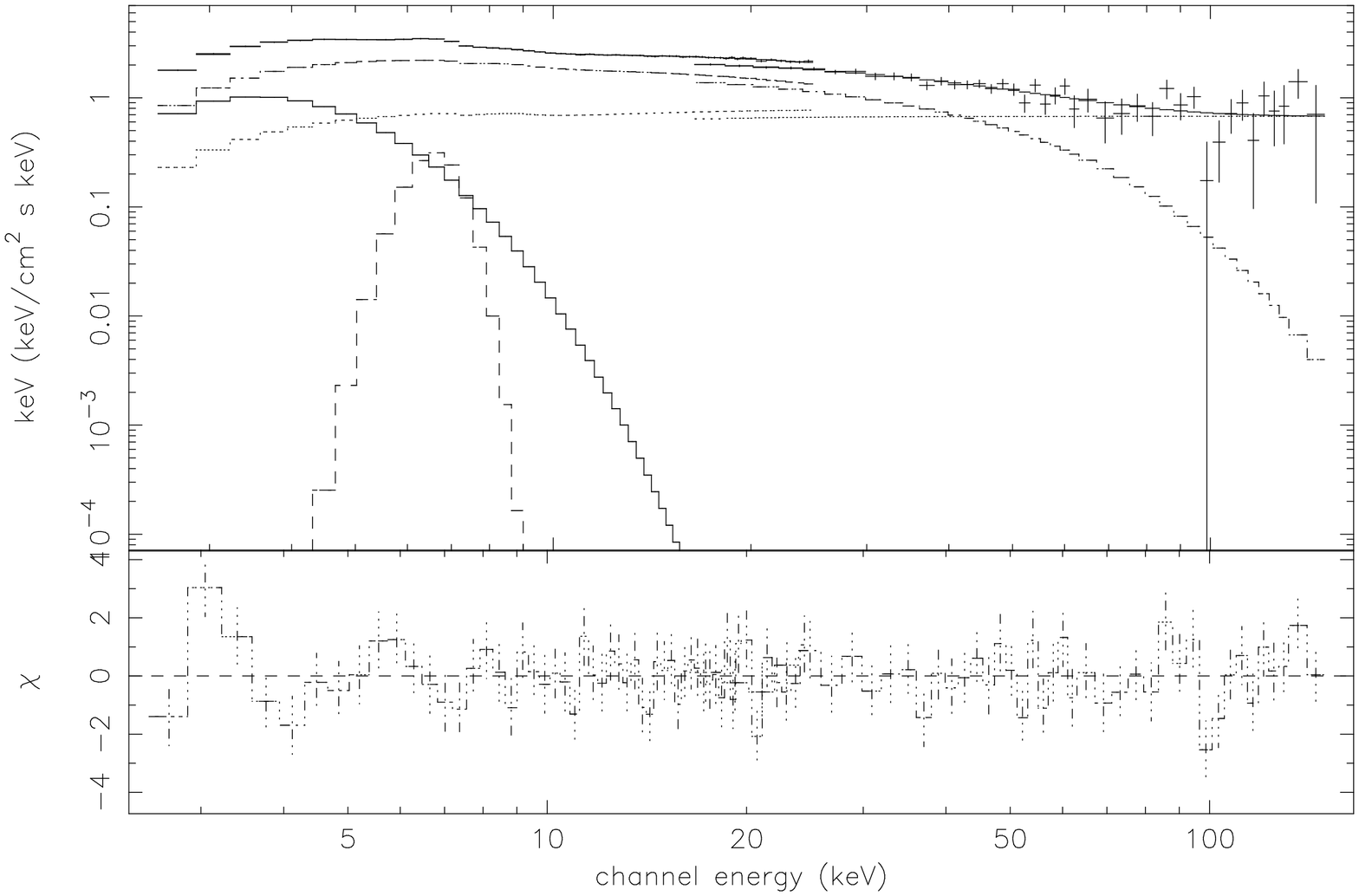} &
\includegraphics[width=2.2in,angle=270]{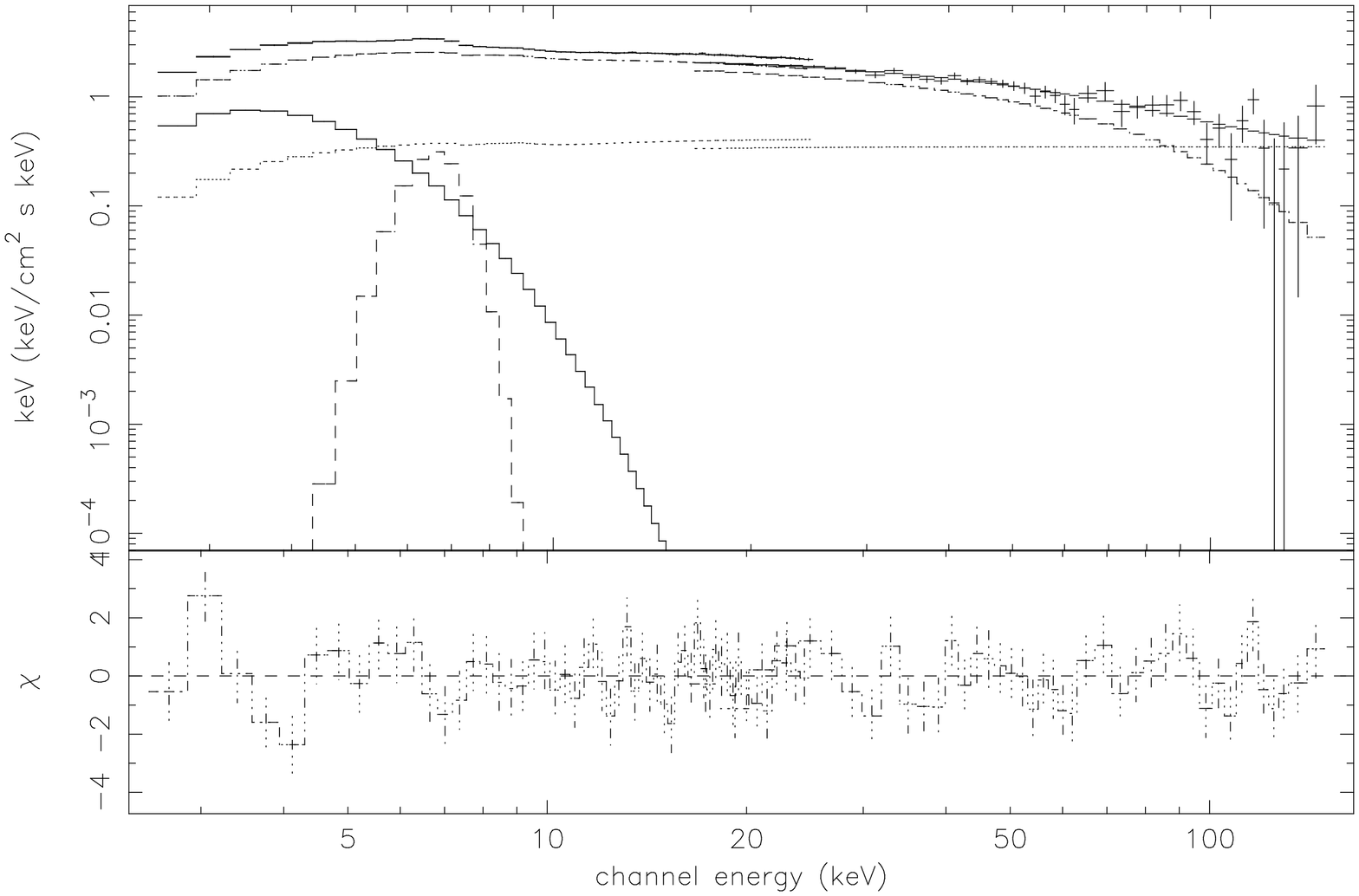} \\
\end{tabular}}
\caption{The unfolded spectrum for the source GRS $1915+015$ of the different parts (left panel A \& right panel B, see Table \ref{tab3}) of a single pointed observation of MJD 50480. \label{fig5}}
\end{figure}}

\clearpage
\newpage
\begin{figure}
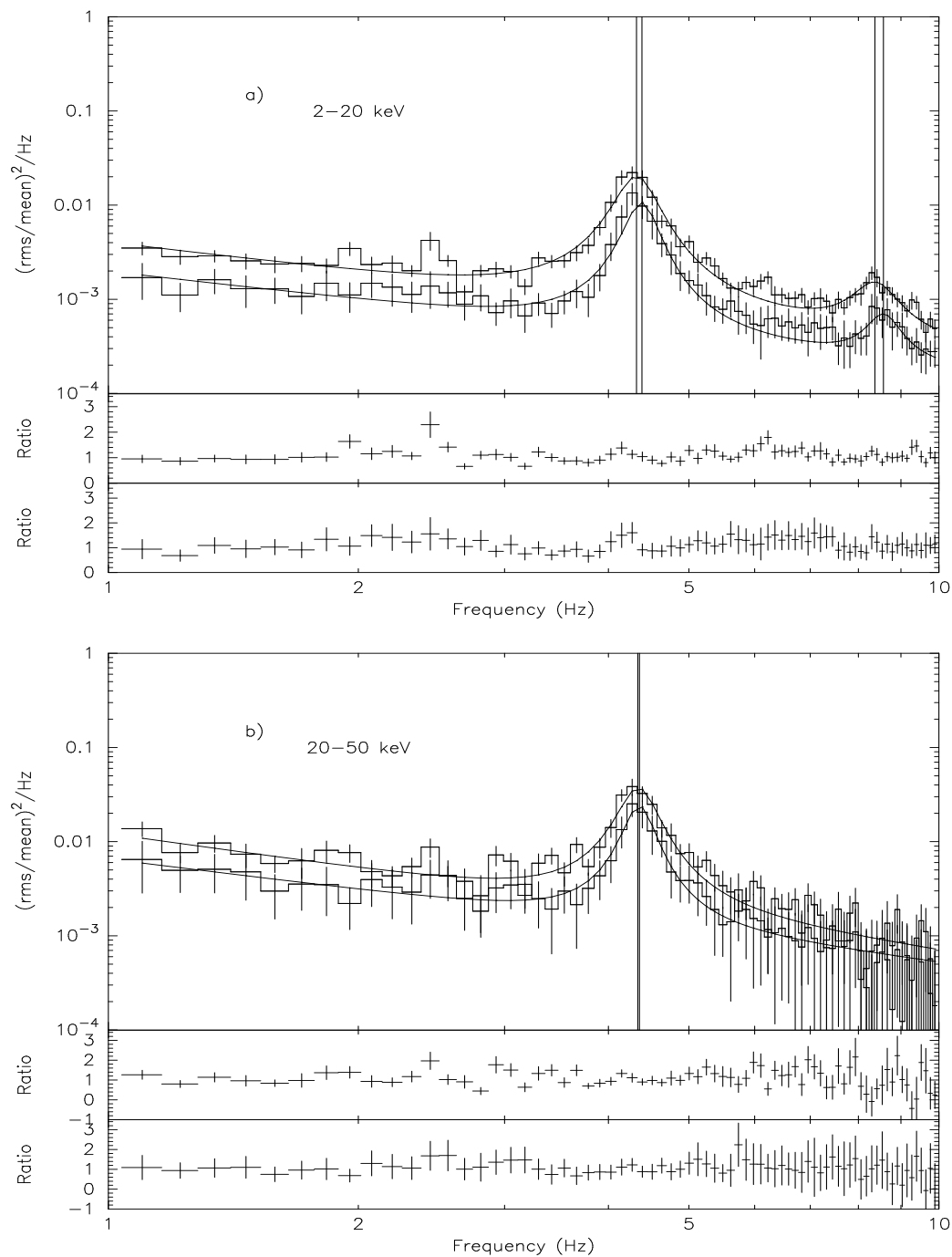

\figurenum{6}{
\begin{tabular}{cc@{\hspace{2pc}}cc}
\includegraphics[width=3.5in,height=5.4in,angle=270]{f6a.eps}\\
\includegraphics[width=3.5in,height=5.4in,angle=270]{f6b.eps} 
\end{tabular}}
\caption{The Power Density Spectrum (PDS) for ~\xtj~  (ObsId:
30191-01-09-00) shown separately for two energy channels (2 - 20 keV and 20 - 50 keV) in (a) and 
(b), respectively. In each figure two parts of the same observations (part A and B; see text) 
are shown, part B data shifted down by a factor of 2 for clarity. Bets fit models
(power law and Lorentzians for QPO) are shown as continuous lines and the residuals
are shown as ratio of data to model at the bottom panels (for part A and B, separately). 
The centroid frequency of QPOs are indicated as vertical lines.\label{fig6}}
\end{figure}

\begin{figure}
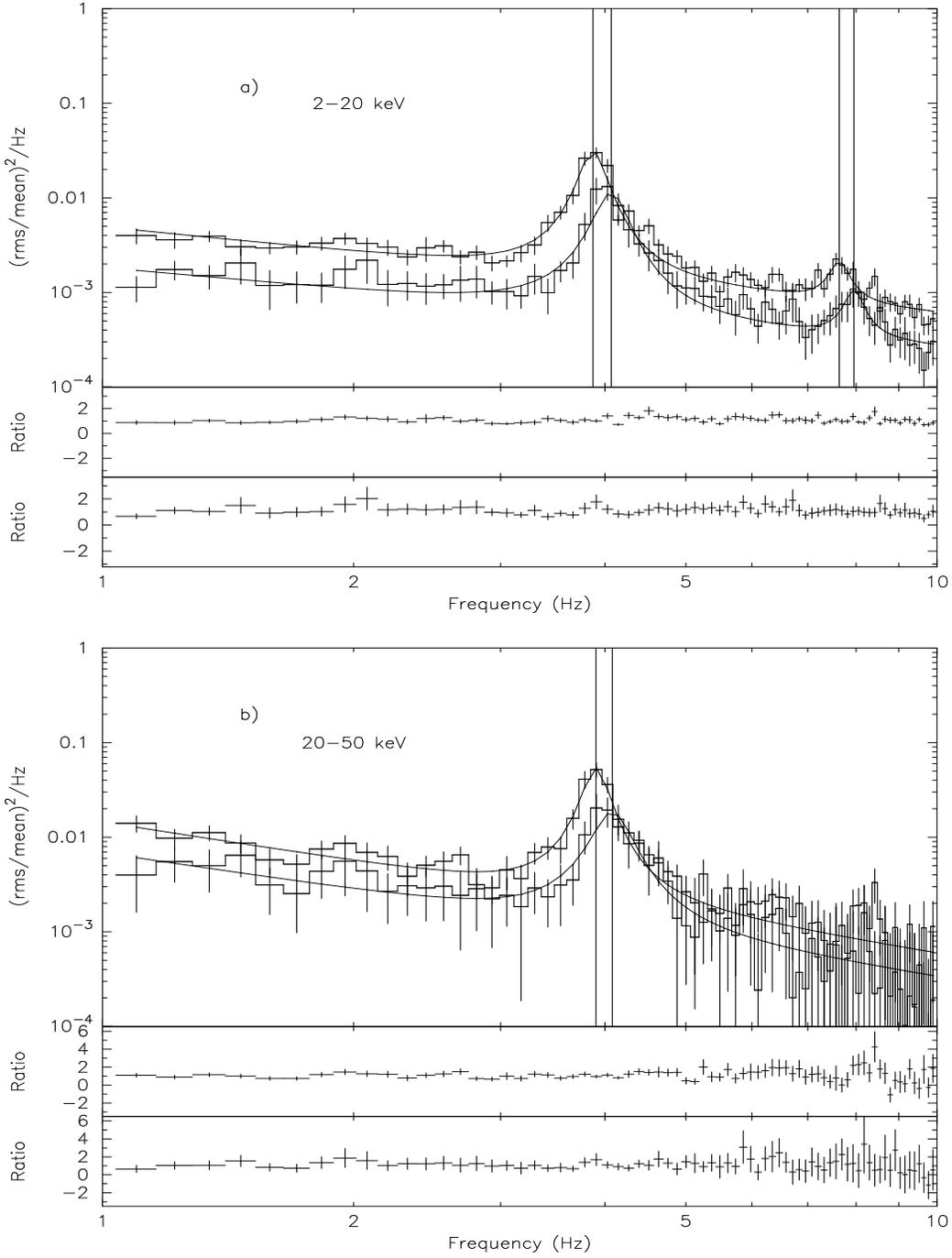

\figurenum{7}{
\begin{tabular}{cc@{\hspace{2pc}}cc}
\includegraphics[width=3.5in,height=5.4in,angle=270]{f7a.eps}\\
\includegraphics[width=3.5in,height=5.4in,angle=270]{f7b.eps}   
\end{tabular}}
\caption{ Same as Figure 6, but for ObsId: 30191-01-09-01. 
\label{fig7}}
\end{figure}

\end{document}